\documentclass[12pt]{article}

\usepackage{graphicx}

\begin{document}

\begin{titlepage}
\begin{flushright}

hep-th/0107232 \\

\end{flushright}

\begin{centering}
\vspace{.41in}
{\large {\bf Inflation in String Theory and how you can get out of
it.}}\\

\vspace{.4in}

 {\bf E.~Papantonopoulos} \\
\vspace{.2in}

 National Technical University of Athens, Physics
Department, Zografou Campus, GR 157 80, Athens, Greece. \\

\vspace{1.0in}

{\bf Abstract} \\

\end{centering}

 Inflation is now a basic ingredient of the modern
cosmology. After reviewing the basic characteristics of inflation,
we briefly discuss inflation in string and brane theories,
focusing on the problem of the exit from inflation in these
theories. We present a model, based on type-0 string theory, in
which our universe after the inflationary phase, passes smoothly
to a flat Robertson-Walker universe.

\vspace{2.3in}
\begin{flushleft}
Talk given at the 2nd Hellenic Cosmology Workshop, National
Observatory of Athens (Penteli), 19-20 April 2001.

\end{flushleft}

\end{titlepage}

\section{Introduction}

Inflation was introduced in cosmology \cite{inflation} in order to
solve three basic problems of standard cosmology. The first
problem is the Particle Horizon. The particle horizon is the
'instantaneous' distance at time t traveled by the light since the
beginning of time and is given by
\begin{equation}
d_{H}(t)=a(t) \int_{0}^{t}\frac{d t^{ \prime}}{a (t^{\prime})}.
\end{equation}
The particle horizon gives the distance at which causal contact
has been established at time t. The present particle horizon is $
d_{H}=2H^{-1}_{0}=6000 h^{-1}Mpc$  where $H_{0}$ is the Hubble
parameter at the present time and $ h $ a parameter with values $(
0.4 \leq h \leq 0.8) $. Recent results from COBE indicate that the
Cosmic Background Radiation (CBR) is uniform in temperature at
large scales. The CBR was emitted at the time of decoupling of
matter and radiation. If one uses the standard cosmology, the
particle horizon at the decoupling time can be calculated and
found $\simeq 0.168 h^{-1} Mpc$. Then if we allow this distance to
expand till the present time, then we will find $\simeq 184 h^{-1}
Mpc$. This means that in the past there was not enough time for
the photons to communicate their temperature to all direction of
the now visible universe.

The particle horizon during inflation (exponential expansion) is
\begin{equation}
d(t)=e^{Ht} \int^{t}_{t_{i}} \frac{d t^{\prime}} {e^{H
t^{\prime}}}
\end{equation}
and grows as fast as $a(t)$. Therefore the successful inflationary
model in which the particle horizon exceeds $2 H^{-1}_{0}$ solves
the Particle Horizon Problem, because the light rays emitted at
the decoupling time, have enough time to communicate the
temperature everywhere and produce a uniform spectrum of CBR, as
it is observed today \cite{maco}.

The second problem of the standard cosmology is known as the
Flatness Problem. The present energy density , $\rho$ , of the
universe has been observed to lie in the relative narrow range
 $0.1 \rho_{c}\leq \rho \leq 2\rho_{c}$, where $\rho_{c}$ is the
critical energy density corresponding to flat universe. Then using
the Einstein equations one finds $\frac{\rho-\rho_{c}}{\rho_{c}}=
3(8 \pi G \rho_{c})^{-1}\frac{k}{a^{2}}$ which is equal to $a$ for
a matter dominated universe. If we use the fact that
$a=t^{\frac{2}{3}}$ in a matter dominated universe, we can
calculate that at the time of decoupling
$\frac{\rho-\rho_{c}}{\rho_{c}}\sim 10^{-16}$. The standard
big-bag cosmology cannot offer a plausible explanation why the
early energy density of the universe is so finely closed to its
critical value. Inflation offers a very convincing argument why
$\rho=\rho_{c}$ at the very early times. Because of the
exponential  growth of the scale factor $a$, the term
$\frac{k}{a^{2}}$ gets suppressed driving
$\frac{\rho-\rho_{c}}{\rho_{c}}\ll 1$ in a natural way. In fact
inflation implies that the present universe is flat to a great
accuracy.

The third problem is the Monopole Problem \cite{monop}. At early
times in the expansion, the physics of the universe is described
by particle theory. These theories predict that at the expansion
procees as the universe cools, phase transitions occur which are
natural consequences of symmetry breaking. During these phase
transitions various particles are produced like magnetic
monopoles. If one calculates the number of monopoles produced
during, for example, the electroweak phase transition, finds that
the dominated matter of our universe should consists out of
monopoles, contradicting with all experimental results. Inflation
solves the Monopole Problem by diluted the primordial monopole
density.

Most of the successful models of inflation incorporate a scalar
field the inflation. The scalar field appears in the action with a
kinetic energy term and a potential term. The interplay between
potential and kinetic energy of the inflaton field gives all the
dynamics in an inflationary model. During inflation, the inflaton
field slowly rolls down the nearly flat potential until it reaches
its minimum. Fluctuations of the inflaton field around the minimum
of the potential reheats the universe, and then involves to its
present epoch.

We all now believe, and recent observational data make this
believe stronger, that the universe in its early evolution had
passed from an inflationary epoch. The crucial question however
is, how long the inflation lasted and how it ended. All the
inflationary models have to solve the three basic problems we
discussed. The solution of these problems impose bounds on the
duration of the inflation. These bounds depend on the parameters
of the theory, but all these bounds give a very short period of
inflation. The end of inflation stars when the inflaton field
rolls down to the minimum of the potential and the reheating
starts. The reheating of the universe is a complex physical
problem \cite{reheat} which depends on the physical parameters of
the theory and in some models a fine tuning of parameters is
needed.

As we already discussed, the inflaton field plays a central
r$\hat{o}$le in inflation. This scalar field, usually is added by
hand in the action, and the resulting effective theory is studied.
Nevertheless, there are models in the literature, mostly based on
Grand Unification Schemes, where the inflaton and its potential
arise naturally from the scalar-field content of the theory.

\section{Inflation in String and High Dimensional Theories}

String theory is a high dimensional theory, which attempts to
unify the electroweak, strong and gravity theories. Cosmological
string models \cite{lidsey} can be obtained if one dimensionally
reduces the string theory to four dimensions and looks for time
dependent solutions in a Robertson-Walker background. The massless
fields that survive this procedure, in the simplest possible
string theory, are the dilaton field, the graviton field and the
action field. The theory that describes such a theory is
\cite{aben}
\begin{equation}\label{action}
\int d^{4}x \sqrt{-g} \left [ R-\frac{1}{2}(D
\phi)^{2}-\frac{1}{2} e^{2\phi}(Db)^{2}-\frac{1}{3} e^{\phi}
\delta c \right ]
\end{equation}
where $\phi$ is the dilaton field, $b$ is the action field and
$\delta c$ is the central charge deficit defined by $c=22+\delta
c$ with $c$ the central charge. Using this action one can write
down the equations of motion (the $\beta$-functions of the theory)
for the graviton, dilaton and action fields. If we allow only time
dependence and choose a Robertson-Walker background, these
equations can be solved to give a linear expanding universe
\cite{aben}.

The action given in (\ref{action}) is the most general action for
any bosonic string theory dimensionally reduced to
four-dimensions. So it is interesting to find that the cosmology
of such a theory is of a linear expanding universe. To get an
inflationary solution we must add a potential for the dilaton
field. Such a term does not arise classically and has to be
generated by string loop effects or by non-perturbative phenomena.

Another attempt to string cosmology was to interpret the dilaton
field of the string theory as the inflaton \cite{venez}, the
scalar field which drives the inflation. The advantage of such the
approach is that the usual four-dimensional Einstein gravity is
naturally part of the theory and you can study the very early
cosmological evolution, evading even the initial singularity. The
disadvantage of such theories is that they invoke on a dilaton
potential for which we know very few things and that it is very
difficult to end the inflationary phase and to go in a smooth way
to the usual Robertson-Walker expanding universe.

We can find an inflationary solution and eventual exit from it in
a string theory, if we consider non-critical strings. In
non-critical strings the $\beta$-functions of the theory (the
classical equations of motion) are modified in such a way as to
restore the conformal invariance \cite{mavrom}. Then the resulting
equations of motion can be solved analytically or nummerically and
study cosmological evolution. In a toy two dimensional model we
had studied such a theory \cite{grace}. The action of this model
is
\begin{equation}\label{action1}
\int d^{2}x \sqrt{-g} \left [ R-2(\nabla \phi)^{2}+2
e^{2\phi}(\nabla T)^{2}+V(T)-e^{2\phi}Q^{2} \right ]
\end{equation}
where $T$ is the tachyon field with a potential $V(T)$ and Q is a
time dependent quantity similar to the central charge of
(\ref{action}). Solving the modified equations of motion,
supplemented by the Curci-Paffuti equation we get an inflationary
phase followed in a smooth way by an expanding Robertson-Walker
universe
 \cite{grace}. In the next section will discuss a four-dimensional
 realistic theory, based on non-critical strings in which the
 inflationary phase is terminated and then followed by the usual
 expanding phase.

 The introduction of branes into cosmology had offered another
 exiting possibility. According to the brane scenario we are
 living on a three dimensional brane, where also the gauge fields
 are confined. The brane is embedded in a higher dimensional space
 in which the gravitational field lives. There are a lot of
 cosmological models in the literature that exhibit
 brane-inflation. We can separate these models into two
 categories. In the first one \cite{statbran}, the brane is
  a static solution of
 the underling theory and the cosmological evolution is due to the
 time development of the fields, while in the second category
 \cite{movbran} the
 cosmological evolution is due to the movement of the brane in the
 bulk.

We had studied \cite{left,papa} a brane moving in a particular
background of a type-0 string background. What we found is that an
energy density (time dependent cosmological constant) was induced
on the brane, because of its movement in the bulk. This energy
density gives an inflationary phase in the cosmological evolution
of the brane-universe which is terminated when the brane is far
way and does not feel the gravitational field of the other branes.

\begin{figure}[h]
\centering
\includegraphics[scale=0.7]{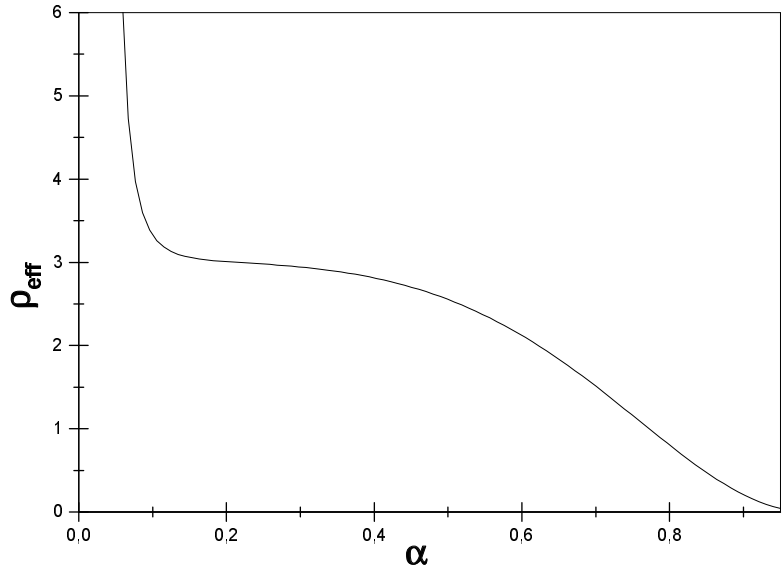}\caption
{The induced energy density on the brane as a function of the
brane scale factor.}
\end{figure}

Motivated by brane cosmology we had studied \cite{pappa} the
four-dimentional Einstein equation with a time dependent
cosmological constant of the form $\Lambda=\frac{1}{logt}$. If $a$
is the scale factor of the universe, the Einstein equation which
gives the cosmological evolution of the $a(t)$ in the presence of
a cosmological constant of the form $\Lambda=\frac{1}{logt}$ is
\begin{equation}\label{common}
\frac{\ddot{\alpha}}{\alpha}=\Big{(}1-\frac{3\gamma}{2}\Big{)}
\Big{(}\frac{\dot{\alpha}^{2}}{\alpha^{2}
}+\frac{k}{\alpha^{2}}\Big{)}+\frac{\gamma}{2}\frac{1}{logt}
\end{equation}
where $\gamma$ appears in the equation of state
$p=(\gamma-1)\rho$. For a flat universe $k=0$, the above equation
has two distinct solutions one of rapid growth (inflation) and the
other of logarithmic growth  (slow expansion). These solutions are
connected in a smooth way, without any singularity.

\section{Cosmological Evolution of a Four-Dimensional String Model
Based on Type-0 String Theory}

We consider a ten-dimensional string theory of type-0
\cite{type0}. We dimensionally reduce the ten-dimensional action
to the four-dimensional space-time on a brane, assuming that all
our fields depend only upon the time. We choose two different
fields to parametrize the internal space. The first field sets the
scale of the fifth dimension, while the other parametrize a
conformally flat five-dimensional space. In the same way, we
dimensionally reduce the modified (because of the
off-criticallity) $\beta$-functions of the ten-dimensional theory,
to the effective four-dimensional $\beta$-functions, assuming a
Robertson-Walker form of the four-dimensional metric. The
resulting equations of motion, supplemented by the equation
Curci-Paffuti, are too complicated to be solved analytically. We
follow a systematic method, of the quasi-linear systems
\cite{book}, to solve them numerically. The action of the theory
in ten dimensions is
\begin{eqnarray}\label{actiontype0}
 S&=&\int d^{10} x
\sqrt{-G}\Big{ [}e^{-2\Phi}\Big{(}R + 4(\partial_M \Phi)^2
-\frac{1}{4}(\partial_M T)^2 \nonumber \\&~~ &
 - \frac{1}{4}m^2T^2 -Q^{2} \Big{)}
 -  \frac{1}{4}(1 + T + \frac{T^2}{2})|{\cal F}_{MNP\Sigma
T}|^2\Big{]}
\end{eqnarray}
where capital Greek letters denote ten-dimensional indices, $\Phi$
is the dilaton, $T$ is the tachyon field of mass $m^2 <0$. In our
analysis we have ignored higher-order terms in the tachyon
potential. The quantity ${\cal F}_{MNP\Sigma T}$ denotes the
appropriate five-form of type-$0$ string theory, which couples to
the tachyon field in the Ramond-Ramond (RR) sector via the
function $f(T)=1 + T + \frac{1}{2}T^2$, and $Q^{2}$ plays the role
of the central charge.

We need first to consider the dimensional reduction of the
ten-dimensional action to the four-dimensional space-time on the
brane. This procedure is achieved by assuming the following ansatz
for the ten-dimensional metric:
\begin{equation}
G_{MN}=\left(\begin{array}{ccc}g^{(4)}_{\mu\nu} \qquad 0 \qquad 0
\\ 0 \qquad e^{2\sigma_1} \qquad 0 \\ 0 \qquad 0 \qquad
e^{2\sigma_2} I_{5\times 5} \end{array}\right) \label{metriccomp}
\end{equation}
where lower-case Greek indices are four-dimensional space time
indices, and $I_{5\times 5}$ denotes the $5\times 5$ unit matrix.
We have chosen two different scales for internal space. The field
$\sigma_{1}$ sets the scale of the fifth dimension, while
$\sigma_{2}$ parametrize a flat five dimensional space. In the
context of cosmological models, we are dealing with here, the
fields $g_{\mu\nu}^{(4)}$, $\sigma_{i},~i=1,2$ are assumed to
depend on the time $t$ only.

Upon considering the fields to be time dependent only, restricting
ourselves to the compactification (\ref{metriccomp}), and assuming
a Robertson-Walker form of the four-dimensional metric, with scale
factor $a(t)$, the modified equations of motion are
\begin{eqnarray}
  &~& \frac{-3\ddot a}{a} + {\ddot \sigma}_1 + 5 {\ddot \sigma}_2 - 2
{\ddot \Phi} + {\dot \sigma}_1^2  + 5 {\dot \sigma}_2^2 +
\frac{1}{4}{\dot T}^2 + e^{-2\sigma _1 + 2\Phi}f_5^2 f(T)=0,
\nonumber \\ &~&  {\ddot a}a + a{\dot a}\left(2Q + {\dot \sigma}_1
+ 5{\dot \sigma}_2 - 2{\dot \Phi}\right)+ e^{-2\sigma_1 +
2\Phi}f_5^2 f(T)a^2 =0~, \nonumber
\\ &~& {\ddot \sigma}_1 + 5{\dot \sigma}_1^2 + 3\frac{{\dot
a}}{a}{\dot \sigma}_1 + 2Q{\dot \sigma}_1 + 5{\dot \sigma}_1{\dot
\sigma}_2 -  2{\dot \sigma}_1{\dot \Phi}+ e^{-2\sigma_1 +
2\Phi}f_5^2~f(T)=0 ~, \nonumber \\ &~&  3{\ddot \sigma}_2 + 9{\dot
\sigma}_2^2 + 3\frac{{\dot a}}{a}{\dot \sigma}_2 + 2Q{\dot
\sigma}_2 + {\dot \sigma}_1{\dot \sigma}_2 -  2{\dot
\sigma}_2{\dot \Phi}- e^{-2\sigma_1 + 2\Phi}f_5^2~f(T)=0~,
\nonumber
\\ &~&  2{\ddot T} + 3\frac{{\dot a}}{a}{\dot T} + Q~{\dot T} +
{\dot \sigma}_1 {\dot T} + 5{\dot \sigma}_2 {\dot T} - 2{\dot
T}{\dot \Phi} + \nonumber \\ &~& m^2T - 4 e^{-2\sigma_1 +
2\Phi}f_5^2 f'(T)=0~, \nonumber \\ &~&  {\ddot \Phi} + Q{\dot \Phi
} + 6\frac{{\dot a}}{a} + 6\frac{{\dot a}^2}{a^2} + \nonumber
\\ &~&2\left[-{\ddot \sigma}_1 - \frac{3{\dot a}}{a}\sigma_1
-5{\ddot \sigma}_2 - 15\frac{{\dot a}}{a}{\dot \sigma}_2 - {\dot
\sigma}_1^2 - 15 {\dot \sigma}_2^2 - 5 {\dot \sigma}_1{\dot
\sigma}_2 - \right.\nonumber \\ &~&2\left.{\dot \Phi}^2 + 2{\ddot
\Phi} + 6\frac{{\dot a}}{a}{\dot \Phi} + 2{\dot \sigma}_1{\dot
\Phi} + 10{\dot \sigma}_2{\dot \Phi} \right] - \frac{1}{4}{\dot
T}^2 + \frac{1}{4}m^2T^2 + Q^2=0~,\nonumber \\ &~&
C_{5}=e^{-\sigma_1 + 5 \sigma_2}f(T)f_5~, \nonumber \\ &~&
{\Phi}^{(3)} + Q{\ddot \Phi} + {\dot Q}{\dot \Phi} + 12\frac{{\dot
a}}{a^3}\left(a{\ddot a} + {\dot a}^2 + Q~a{\dot a}\right) -{\dot
T}({\ddot T}+ Q{\dot T}) + \nonumber \\ &~& 4{\dot
\sigma}_1({\ddot \sigma}_1 + 2{\dot \sigma}_1^2 + Q{\dot
\sigma}_1) + 20{\dot \sigma}_2({\ddot \sigma}_2 + 2{\dot
\sigma}_2^2 + Q{\dot \sigma}_2)=0 \nonumber
\end{eqnarray}
where $f'(T)$ denotes functional differentiation with respect to
the field $T$, the dots denote time derivatives, $\Phi^{(3)}$
denotes triple time derivative.

To solve the above system numerically, we separate the fields in
their asymptotic values plus fields which tend asymptotically to
zero. Substituting these fields back to the equations we let the
system evolve in time backwards \cite{joint}.

Our main results are as follows. The cosmological evolution of our
universe passes through the following phases. At very early times,
the universe starts from the initial singularity. Then enters a
phase where the physical dimensions are formed. At this stage
$\sigma_{1}$, $\sigma_{2}$ and $a$ are comparable in magnitude.
The phase of inflation follows, during which the scale factor
grows exponentially (for a short time though), while the internal
space contracts with very negative values of $\ddot{\sigma_{1}}$
and $\ddot{\sigma_{2}}$. Finally the universe enters a phase,
where it expands slowing until it reaches the asymptotically flat
space. The internal space continuous to contract but with very
slow rate (with $\ddot{\sigma_{1}}$ and $\ddot{\sigma_{2}}$
positive) until it reaches a constant value. The fields
$\sigma_{1}$ and $\sigma_{2}$ scale differently. The field
$\sigma_{2}$ very soon freezes to a value much lower than the
value of $\sigma_{1}$, indicating the the fifth dimension can be
much larger than the other five dimensions. In Fig.2 the evolution
of the scale factor is shown.

\begin{figure}[h1]
\centering
\includegraphics[scale=0.9]{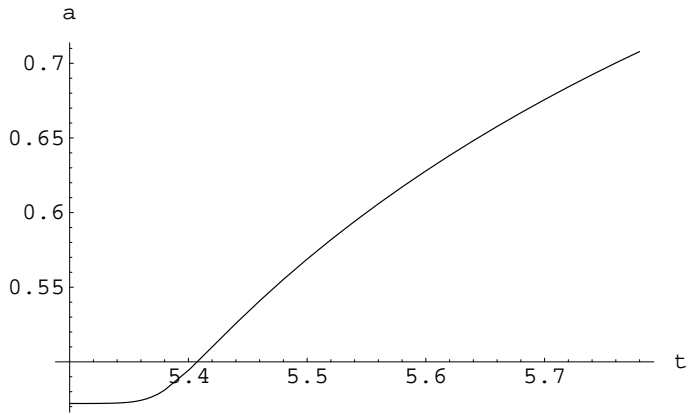}
\caption {The evolution of the scale factor in time.}
\end{figure}

The dilaton field at the singularity is infinite, indicating that
the gravity is very strong. Then at the second phase of evolution,
the strength of gravity is weakened because the dilaton field
drops linearly. Then during inflation the gravity becomes more
weaker, and finally at the exit, the dilaton field continuous to
drop linearly. The tachyon field  during the evolution, falls
continuously until it becomes zero. The RR-field, from zero value
at the singularity, grows until it reaches a constant value at the
exit of inflation.

\section{Summary}

We have presented a cosmological model based on a type-0 string
theory. The type-0 string theory is rich in its content. Except
the graviton and the dilaton field, it includes also a tachyon
field which couples to an RR five-form field. To avoid tachyon
instabilities the tachyon field has to take values at the minimum
of its potential. Considering the ten-dimensional action of the
type-0 field theory one can get at the conformal point the
$\beta$-functions of the theory. Reducing them to four-dimensions
and assuming that all the fields are time dependent we get an
effective four-dimensional theory. We found that this theory in a
Robertson-Walker background has an inflationary phase but we
cannot exit from this phase in a smooth way \cite{joint}.

We believe that the central issue of inflationary cosmology is not
how we can get an inflationary phase, but how we can exit from it.
In view of the recent discussion \cite{eternal} of the exit from
inflation in string theory, and in general on the consistent
quantization of de-Sitter Universes \cite{Witten}, we think that
it is useful to propose new mechanisms to exit from inflation. In
our previous work \cite{grace}, we had proposed a mechanism based
on non-critical strings, to exit from inflation. We had consider a
two dimensional model with a tachyon field. The presence of the
tachyon field was crucial for an inflationary phase to go smoothly
to a Robertson-Walker phase.

In this work we applied this mechanism to a realistic
four-dimensional model. We modified the $\beta$-functions of the
ten-dimensional theory, assuming that the string theory is
non-critical. This hypothesis, introduces new terms in the
$\beta$-functions of the theory. Physically these terms express
the fact that our non-critical string theory is performing small
oscillations around the conformal point.

The modified $\beta$-functions supplamented with the Curci-Paffuti
equation, are then reduced to four-dimensions. Assuming an
homogeneous and spherically symmetric background, we solve
nummerically the resulting equations. The numerical analysis of
the system of the coupled differential equations uses the
quassi-linear method. Because we wand the solution asymptotically
to approach the Minkowski space with a linear dilaton, we separate
the fields in their asymptotic values plus fields which tend
asymptotically to zero. Substituting these fields back to the
equations we leave the system to evolve in time backwards. What we
find is that the scale factor after the initial singularity,
enters a short inflationary phase and then in a smooth way goes to
flat Minkowski space. The fields $\sigma_{1}$ and $\sigma_{2}$
which parametrise the internal space have an interesting
behaviour. The field $\sigma_{1}$ which sets the scale of the
fifth dimension, during inflation contracts until it reaches a
constant value. After the inflation maintains this value until the
universe involves to an asymptotically flat space. The field
$\sigma_{2}$ which parametrise the flat five-dimensional space
freezes to a constant value which is much smaller than the value
of the constant of the fifth dimension. Thus we see that
cosmological evolution may lead to different scales if we
decompose the extra dimensions. It is important finally to notice,
that this difference in scales of the extra dimensions is due to
the fact that in our theory the gravity is very weak
asymptotically.

\section*{Acknowledgments}

 This Talk is based on work done
 in collaboration with G. Diamandis, B. Georgalas, N. Mavromatos
 and I. Pappa. We thank the
organizers of the 2nd Hellenic Cosmology Workshop, National
Observatory of Athens (Penteli), 19-20 April 2001, for their kind
invitation.

\end{document}